\documentclass[conference, a4paper]{IEEEtran}
\IEEEoverridecommandlockouts
\usepackage{cite}
\usepackage{amsmath,amssymb,amsfonts}
\usepackage{multirow}
\usepackage{algorithmic}
\usepackage{graphicx}
\usepackage{textcomp}
\usepackage{xcolor}
\usepackage{color, colortbl}
\def\BibTeX{{\rm B\kern-.05em{\sc i\kern-.025em b}\kern-.08em
    T\kern-.1667em\lower.7ex\hbox{E}\kern-.125emX}}
\begin{document}

\title{Towards Big Data Modeling and Management Systems: From \textbf{DB}MS to \textbf{BD}MS}

\author{\IEEEauthorblockN{Rania Mkhinini Gahar}
\IEEEauthorblockA{\textit{OASIS Laboratory} \\
\textit{National Engineering School}\\
\textit{of Tunis,}\\
\textit{University of Tunis El Manar}\\
Tunis, Tunisia \\
rania.mkhininigahar@enit.rnu.tn}
\and
\IEEEauthorblockN{Olfa Arfaoui}
\IEEEauthorblockA{\textit{RISC Laboratory} \\
\textit{National Engineering School}\\
\textit{of Tunis,}\\
\textit{University of Tunis El Manar}\\
Tunis, Tunisia \\
olfa.arfaoui@isln.u-carthage.tn}
\and
\IEEEauthorblockN{Minyar Sassi Hidri}
\IEEEauthorblockA{\textit{Computer Department} \\
\textit{Deanship of Preparatory Year }\\
\textit{and Supporting Studies,}\\
\textit{Imam Abdulrahman Bin Faisal University}\\
Dammam, Saudi Arabia \\
mmsassi@iau.edu.sa}
}
\maketitle

\begin{abstract}
To succeed in a Big Data strategy, you have to arm yourself with a wide range of data skills and best practices. This strategy can result in an impressive asset that can streamline operational costs, reduce time to market, and enable the creation of new products. However, several Big Data challenges may take place in enterprises when it comes to moving initiatives of boardroom discussions to effective practices. From a broader perspective, we take on this paper two very
important challenges, namely modeling, and management. The
main context here is to highlight the importance of understanding
data modeling and knowing how to process complex data while supporting the characteristics of each model.
\end{abstract}

\section{Introduction}
In today's society, data is growing exponentially. It therefore becomes more complicated to manage these with traditional tools. The \textit{Big Data Analytics} process was therefore created to manage this mass of data and draw results from it.

Where staggering amount of data is meaningful, Big Data faces colossal challenges which are good for mastering in order to keep this database under control. Stored in multiple Data Centers, the exploitation of Big Data continues to grow, especially with the popularization of Cloud Computing (remote and online storage system) \cite{naeem2022trends,pramanik2023analysis,GaharAHH19,AlsaifHFEH22,AlsaifHH21,HidriZA18,ZoghlamiHA16}.

Information processing is one of the main challenges of Big Data. Indeed, data arrives in droves and in all formats from the four corners of the world, at all times. Companies in charge of Data Centers must therefore set up management tools capable of monitoring the velocity of data. At the same time, the quality and relevance of the information received must also be checked.

In this context, data modeling and management are two of the most important and valuable tools for understanding business information. The Big Data modeling concept implied two terminologies which are "Data modeling" and "Big Data". The "Big Data" term means all digital data produced by the use of new technologies for personal or professional purposes. This data kind  is complex by nature too. That's why it is impossible to be analyzed using traditional methods \cite{ZoghlamiHA15,arAHH19}.

Data with such complexity can be analyzed using high-quality data modeling methods. In this context, it should be clear that the \textit{Data modeling} includes the organizing data method in such visualized patterns that the data analysis process can be performed with aptitude. These techniques include the process of making visual representations of the whole or part of the datasets \cite{shi2022advances}.

Thereby, it employs a certain data modeling method. That's why it is different from the traditional methods and process consists to organize Big Data for the companies' use.

Whereas Big Data management is  a sort of organization, administration, as well as governance of both large volumes namely structured and unstructured data. The Big Data management target is concluded in a high data quality level and accessibility for business intelligence and Big Data analytics applications. Many organizations such as corporations, government agencies, and others adopt Big Data management strategies to lead them contending with fast-growing data pools, typically involving many terabytes or even petabytes stored in several file formats variety. 

Effective Big Data management can help companies to locate valuable information in large unstructured and semi-structured datasets from various sources, including call detail records, system logs, sensors, images, and social media sites.

The remainder of this paper consists of two sections. Big Data Modeling is highlighted in Section II. Some Big Data Management systems are presented and subsequently described in Section III. 
The overall conclusion with future extension
remarks are stated in section IV.

\section{Big Data Modeling}
\begin{figure}[htbp]
\centerline{\includegraphics[width=2.3cm]{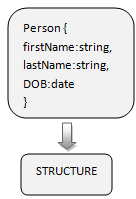}}
\caption{The Person structure.}
\label{fig1}
\end{figure}

The Big Data modeling concept depends on many
factors. It includes the data structure, the operations that can be performed on the applied ones, and constraints to models \cite{bachechi2022big}. It is necessary to determine the data characteristics before it can be manipulated or analyzed in a meaningful as well as significant way \cite{ZoghlamiHA15,Souli-JbaliH15}. Let's take for example the structure \textit{Person} whose characteristics are resumed in surname, the first name, and the Date Of Birth (DOB) as shown in Fig. \ref{fig1}.

\begin{figure}[htbp]
\centerline{\includegraphics[width=2.2cm]{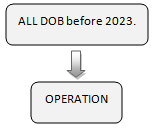}}
\caption{An operation example.}
\label{fig2}
\end{figure}

Likewise, the fact that we can perform data arithmetic or aggregation with the DOB field, and not the first name field which is categorical, is also part of our understanding of the data model.
These are nothing but operations that can be performed. Let us cite the example of the selecting operation of all persons having DOB before 2023 as described in Fig. \ref{fig2}.

\begin{figure}[htbp]
\centerline{\includegraphics[width=3.7cm]{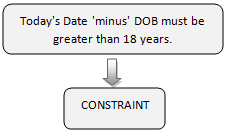}}
\caption{A constraint example.}
\label{fig53}
\end{figure}

Finally, we can know that in this society, the age
corresponding to the current date minus the DOB cannot be under 18 years old. A translation of this constraint can be given by Fig. \ref{fig53}. 

So this overviews a way to detect records with obviously wrong DOB.

\subsection{Data Models types}
\subsubsection{Relational data model}
It refers to a way of structuring information in a matrices form called tables or relations. This very simple model is by far the most widespread in Database Management Systems (DBMS), which are thus called relational DBMSs \cite{Benali-SouguiHG16,SassiGOA10}. A relational database, therefore, consists of a structured dataset in the form of relations. It is similar to the table in Table. \ref{tab1fg} presented here for an employee application.
However, we should pay attention to relational tables, called relationships. This array actually represents a tuple set. In Table. \ref{tab1fg}, relational tuple is framed in red. It is represented by a row in the table. A relational tuple implies that, unless otherwise specified, its elements such as 203 or 204, Mary, etc., are \textit{atomic}. 
\begin{center}
\begin{table}[ht!]
\caption{Employee table.}
\begin{tabular}{c}
  \includegraphics[width=9cm]{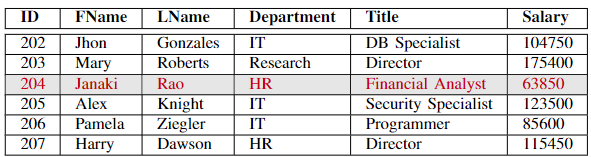}	
\end{tabular}
\label{tab1fg}
\end{table}
\end{center}
The previous example describes a set of six tuples also called records. In fact, when we talk about a collection of \textit{distinct elements} of the same type, it means that it will be impossible to add a tuple that already exists to the solution. By that, if we do, it will be a \textit{duplicate} (see Table. \ref{fighhfg}).

\begin{center}
\begin{table}[ht!]
\caption{Employee table with duplicate.}
\begin{tabular}{c}
  \includegraphics[width=9cm]{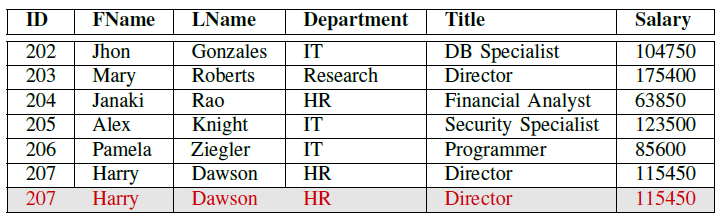}	
\end{tabular}
\label{fighhfg}
\end{table}
\end{center}

Table. \ref{tabbbb} shows another tuple that cannot be added. The latter has all the right attributes, but unfortunately, all are placed in the wrong order. In this way, we called this tuple a \textbf{dissimilar} one.
\begin{center}
\begin{table}[ht!]
\caption{Employee table with dissimilar tuple.}
\begin{tabular}{c}
  \includegraphics[width=9cm]{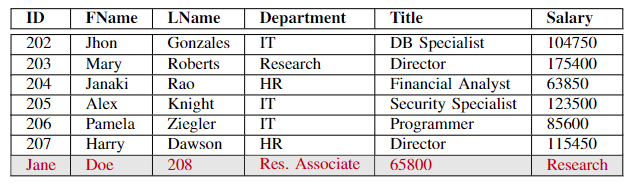}	
\end{tabular}
\label{tabbbb}
\end{table}
\end{center}
The question that arises here is how does the system know that this tuple is different? This draws our attention to the first line which is Table.  \ref{table4}. It is a part of the table schema and simply gives us information about the table name, in this case, \textit{Employee}.

\begin{table*}
 \caption{Employee table with dissimilar tuple.}
 \begin{center}
\begin{tabular}{c}
 

\includegraphics[width=12cm]{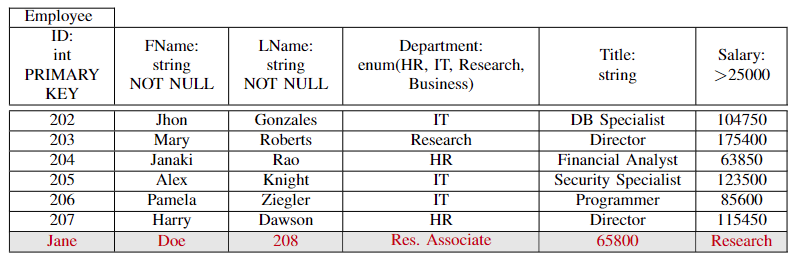}	

\label{table4}
\end{tabular}
\end{center}
\end{table*}

It presents clearly the names of the columns which we  called also attributes relationship. Each column describes its specific data type, i.e. the type constraint for each column. Given this schema, we now need to understand why the last red row does not belong to this table. The schema in a relational table can also specify constraints.

Let us introduce a new table containing employee salary history. Employees are identified with the \textit{EmpID} column, but these are not new values for this table. These are the same IDs present in the ID column of the Employee table, presented previously. This is reflected in the statement made to the right. 

\begin{figure}[htbp]
\centerline{\includegraphics[width=7cm]{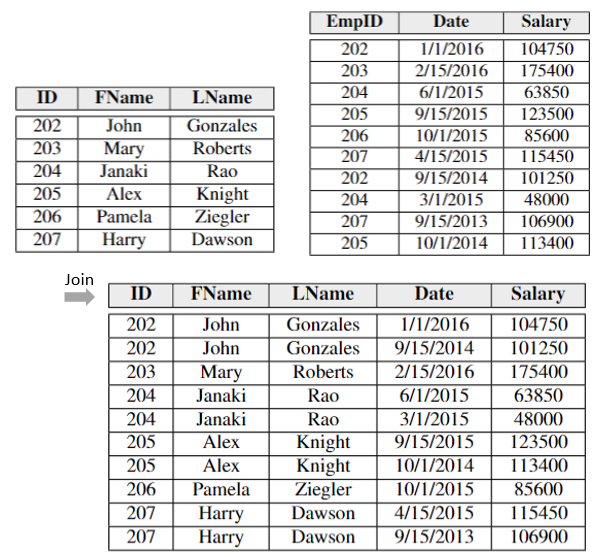}}
\caption{Join relation.}
\label{figjoin}
\end{figure}

References mean that values in one column can only exist if the same values exist in the \textit{Employee} table (see Fig. \ref{figjoin}), called \textit{parent table}. That's why, in relational model concept, the \textit{EmpID} column of the \textit{EmpSalaries} table is called a foreign key which does refer to the primary key of the \textit{Employee} table (see Fig. \ref{figjoin}).

\subsubsection{Semi-structured data model}
Semi-structured data is an intermediate form. They are not organized according to a complex method that makes sophisticated access and analysis possible; however, certain information may be associated with them, such as metadata tags, which allow the addressing of the elements they contain. For example, a Word document is generally considered to be a collection of unstructured data. However, you can add metadata to it in the form of keywords that represent the content of the document and make it easier to find when searching for those terms \cite{hamouda2019semi}. The data is then semi-structured.
\subsubsection{Non-Structured data model}
Unstructured data is defined as data present in absolute raw form. This data is difficult to process due to its complex organization and formatting. Unstructured data management can take data in many forms, including social media posts, chats, satellite imagery, IoT (Internet of Things) sensor data, emails, and presentations, to organize it in the logical and predefined way in data storage. In contrast, the meaning of structured data is data that follows predefined data patterns and is easy to analyze. Examples of structured data would include alphabetized customer names and properly organized credit card numbers \cite{eberendu2016unstructured}.

Unstructured data can be anything that is not in a specific format. It can be a paragraph from a book with relevant information or a web page. An example of unstructured data could also be log files that are not easily separated. Comments and publications on social networks must be analyzed \cite{fadili2016towards}.

\section{Big Data Management}
The data management system refers to the set of practices necessary for the construction and maintenance of a framework for the data import, storage, exploration, and archiving that are necessary for business activities. Data management is the backbone that connects the different segments of the data life cycle in the company \cite{luazuaroiu2022deep}. Data management works hand-in-hand with the management process to ensure that different teams take the necessary steps to always have the cleanest and most up-to-date data. In other words, it is the process to manage that your employees are empowered to monitor changes and trends in real-time.

For example, each data access task, such as finding employees in a department sorted by salary or finding employees in all departments sorted by start date, must be translated by a program according to the request requested. To do this, each request is associated with a developed program to respond to it even for accessing data or updating it.

The third problem concerns constraints. Data types are a way to restrict the nature of data that can be stored in a table. For many applications, however, the constraint provided by this bias is too coarse. For example, a column that contains the price of a product should only accept positive values. But there is no standard data type that only accepts positive values. Another problem can arise from wanting to constrain the data in one column relative to other columns or rows. For example, in a table containing product information, there can only be one row per product number.

For this, the Structured Query Language (SQL) allows you to define constraints on columns and tables. Constraints give as much control over table data as a user wants. If a user attempts to store data in a column in violation of a constraint, an error is thrown. This applies even if the value comes from the default value definition. Many constraints are called for integrity. For example, say that each employee has exactly one job title \cite{rossi2022characterizing,SassiGOA10}.

Atomicity means that database updates must be "atomic", i.e. they must be done completely or not at all. Out of 5000 rows to be modified, if one modification just failed, then the entire transaction must be rolled back. It is important to note that each modified row can be affected by the modification context of the adjacent one, and any break in that context can have disastrous consequences \cite{colombo2019access, cantabella2019analysis,SouguiHG14}.

When it comes to Big Data, things change. It is clear that traditional DBMS will not deal with massive characteristics. That's why another concept is born. It is baptized BDMS for Big Data Management Systems.
\paragraph{Redis - An Enhanced Key-Value Store} it is called an in-memory data structure store
(in-memory): It can keep data on disks and
saves its state. However, it is intended to make optimal use of memory and memory-based methods to make a number of common data structures very fast for many users \cite{matallah2020evaluation}.
   Redis supports a list of data structures namely: strings, hashes, lists, sets, sorted sets
   \begin{itemize}
       \item  Look-up problem: Now, in the simplest case, a search requires a key-value pair where the key is a string and the value is also a string. So, for a search, we provide the key and get the value and it is simple.
       \item Partitioning and replication:  they are techniques that build the foundation of using Redis as a distributed system. They will be examined as very basic building blocks. For more complex needs, there are more complex abstractions, like Redis Sentinel and Redis Cluster, that build upon these building blocks. Fig. \ref{fig455} describes an example of the Master/Slave replication mode.
       
       \begin{figure}[ht!]
\centerline{\includegraphics[width=8cm]{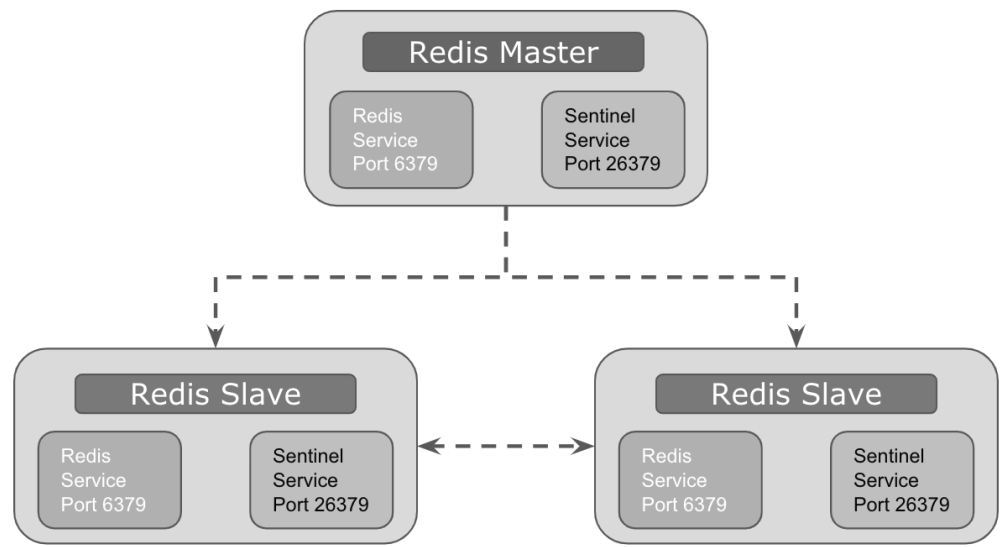}}
\caption{Master/slave replication mode under Redis.}
\label{fig455}
\end{figure}
   \end{itemize}
   
\begin{figure}[ht!]
\centerline{\includegraphics[width=8cm]{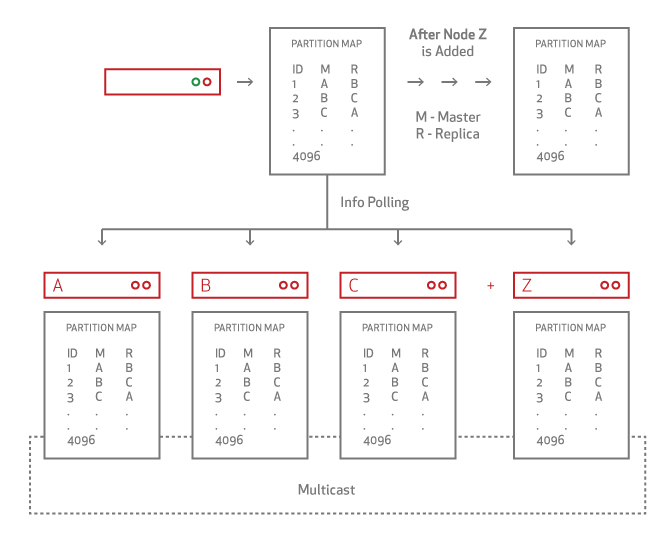}}
\caption{Aerospike architecture.}
\label{fighh5}
\end{figure}

\paragraph{Aerospike: A New Generation KV Store} is an open-source In-Memory Not oly SQL (NoSQL) DBMS. It is a key-value base designed to provide sub-millisecond response times to applications \cite{srinivasan2016aerospike}. 
Fig. \ref{fighh5} can further describe its architecture.

The upper layer presents several applications for real-time systems for consumers, such as travel recommendation systems, pricing engines used for stock market applications, real-time decision systems that analyze data to determine whether an investment must be made, etc.

Nowadays, all data management systems have a common need which resides in the accessibility at any time to the colossal volume of data. The Aerospike system can interact with systems based
Hadoop, Spark, a Legacy database, or even with a
real-time data source. It can exchange large volumes of data with any of these sources and serve quick queries and searches to the above applications. Now, this translates to very high availability and robust consistency requirements.

The storage layer uses three types of storage systems, in-memory with dynamic Random Access Memory (RAM) or Dynamic RAM (DRAM), disk in normal rotation, and a flash disk / Solid-State Drive (SSD), which is a device solid-state for fast data loading when needed. In fact, the Aerospike system has optimized its performance keeping in mind the characteristics of an SSD drive. For those who don't know what an SSD is, you can consider as a kind of storage device whose random read performance is much faster
than that of a hard disk and write performance is
a little slower.

\paragraph{AsterixDB: A DBMS of Semistructured Data} is a shared-nothing parallel DBMS that is used to split data among various nodes by involving a hash-based partitioning mechanism. It also provides a platform for applications that are characterized by scalable storage and analysis of very large volumes of semi-structured data.

\begin{figure}[ht!]
\centerline{\includegraphics[width=7.5cm]{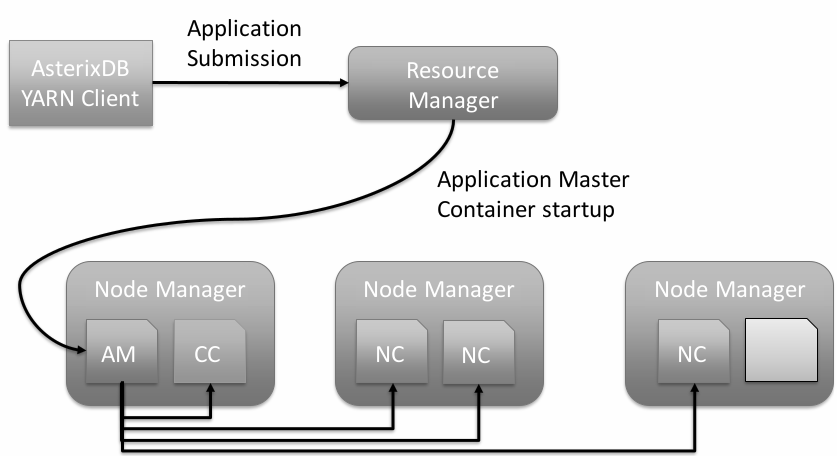}}
\caption{Illustration of a simple YARN cluster with AsterixDB processes and their locations.}
\label{figgg5}
\end{figure}

Fig. \ref{figgg5} provides an overview of how the various software components of AsterixDB map to nodes in a shared-nothing cluster, what is called Asterix Manager (AM) interface. It is composed of three Node Controllers (NCs) and one Cluster Controller (CC). The topmost layer of AsterixDB is a parallel DBMS, with a full, flexible AsterixDB Data Model (ADM) and AsterixDB Query Language (AQL) for describing, querying, and analyzing data. ADM and AQL support both native storage and indexing of data as well as analysis of external data (e.g., data in Hadoop Distributed File System(HDFS)).

In AsterixDB, data is stored in datasets. Each record conforms to the datatype associated with the dataset. In fact, data is hash-partitioned (primary key) across a node set which forms the node group for a dataset and defaults to all nodes in an AsterixDB cluster \cite{alsubaiee2014asterixdb}.
\paragraph{Solr - Managing Text} is a powerful search engine, based on Apache Lucene, integrated with Hadoop. It computes the Term Frequency (TF) and Inverse Document Frequency (IDF) of the collection. Term Frequency-Inverse Document Frequency (TF-IDF) term vectors are often used to represent text documents when performing text mining and machine learning operations.

Practically, other calculated numbers or properties associated with the terms will also be included in the index \cite{shahi2015apache}.

\begin{figure}[htbp]
\centerline{\includegraphics[width=5.5cm]{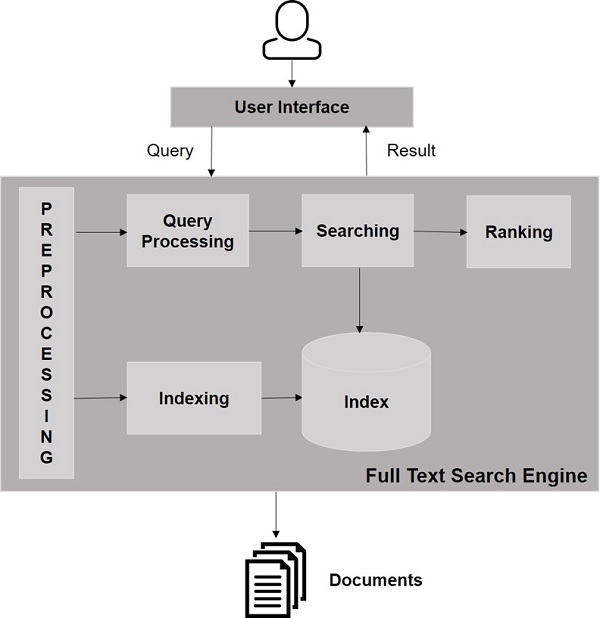}}
\caption{Apache Solr architecture.}
\label{ddfig5}
\end{figure}

The main Solr features are multiple such as indexing of text Document (DOC), Portable Document Format (PDF), PowerPoint (PPT), or Microsoft Excel spreadsheet (XLS) documents,
indexing a database or even the ability to do advanced searches. These are full-text indexes where text columns are supplemented with indexes for other data types, including numeric data, dates, geographic coordinates, and fields where domains are limited to a set of emerging values. Fig. \ref{ddfig5} shows its architecture.
\paragraph{Vertica - A Columnar DBMS} is a relational analytical database that integrates with SQL solutions and Hadoop, Spark, or Kafka architectures, whether in the Cloud (Google, Amazon Web Service (AWS), Azure) or On-Premise. Its performance, scalability, and native high availability allow both startups and the largest global players to carry out their Business Intelligence(BI) or Data Science projects regardless of the volume handled \cite{lamb2012vertica}. Vertica has advanced analytical functions and Machine Learning algorithms to perform part of the in-database processing. It is a columnar data storage platform designed to handle huge volumes of data. This allows its users fast and efficient query performance while providing high availability and scalability on enterprise servers. The main features of the Vertica database are:
\begin{itemize}
\item Column-based storage organization;
\item SQL interface with integrated analysis capabilities;
\item Compression to reduce storage costs;
\item Compatible with programming interfaces;
\item High performance and parallel data transfer.
\end{itemize}

\begin{figure}[htbp]
\centerline{\includegraphics[width=7.2cm]{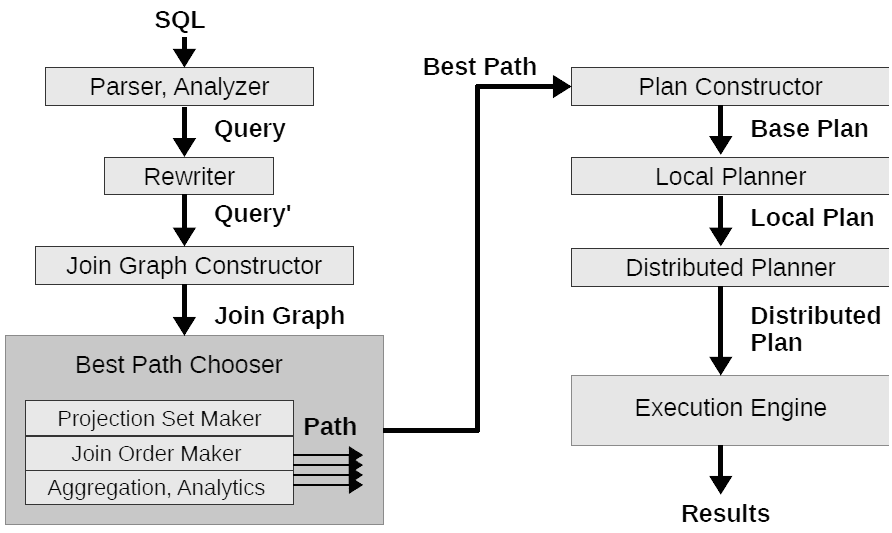}}
\caption{Vertica query example.}
\label{fi65}
\end{figure}

For the query example shown in Fig. \ref{fi65}, a column store reads only three columns while a row store reads all columns.

Table \ref{tabcomp} presents a comparative study of the different BDMS already described above.

\begin{table*}
\caption{Redis vs Aerospike vs AsterixDB vs Solr vs Vertica.}
\begin{center}
    \scriptsize{
\begin{tabular}{|l|p{2.7cm}|p{2.9cm}|p{2.3cm}|p{1.9cm}|p{2.3cm}|}
\hline
\textbf{Name } & \textbf{Redis} &\textbf{Aerospike} & \textbf{AsterixDB}&\textbf{Solr}&\textbf{Vertica}\\ \hline\hline
\textbf{Primary database model}  & Key-Value store &Document store, Key-Value store, Spatial DBMS. &Textural, temporal, spatial& Search engine      & Relational DBMS.                   \\ \hline
\textbf{Implementation language} & C & C & Java& Java               & C++                                         \\ \hline
\textbf{Data scheme}             & schema-free & Schema-free& Both schema-less and schema-full                           & Yes                & Semi-structured / Unstructured, complex hierarchical, and/or queried.\\ \hline
\textbf{Partitioning methods}    & Sharding& Sharding    & Hash-based, partitioning.& Sharding           & Horizontal partitioning, hierarchical.\\ \hline
\textbf{Replication methods}     & Multi-source replication, Source-replica replication.& Selectable replication factor& Primary and remote, replicas.                           & Yes                & Multi-source replication.\\ \hline
\textbf{Transaction concepts}    & Atomic execution of command, blocks and scripts and optimistic, locking.& Atomic execution of operations.                & Atomicity, Consistency, Isolation and Durability, ACID. & Optimistic locking. & ACID\\ \hline
\end{tabular}
}
\end{center}
\label{tabcomp}
\end{table*}

\section{Conclusion}
Data modeling as well as management are very important tasks nowadays for the data scientist. the main reason for recourse is decision-making. In fact, data modeling is the process that make companies able to discover, design, visualize, as well as standardize and even deploy high-quality data assets through an intuitive graphical interface. A proper data model can now serve as a blueprint for designing and deploying databases, leveraging higher quality data sources to improve the application development process and make better decisions \cite{ribeiro2015data,patel2019effective}. Thus, among other things, data Visualisation represents also a challenge that we can't ignore \cite{mohammed2022big}. However, conventional visualization techniques cannot handle the enormous volume, variety, and velocity of data. To do this, several tools have emerged and are constantly evolving. So, we will be interested in Big Data visualization.

\bibliography{bibRania1}
\bibliographystyle{IEEEtran}
\end{document}